\documentclass{ws-mpla}
\usepackage{hyperref}
\usepackage{cite}

\begin{document}

\def\pt{\ensuremath{p_{\mathrm{T}}}}
\def\antibar#1{\ensuremath{#1\bar{#1}}}
\def\tbar{\ensuremath{\bar{t}}}
\def\ttbar{\antibar{t}}
\def\met{\ensuremath{E_{\mathrm{T}}^{\mathrm{miss}}}}
\def\ifb{\mbox{fb$^{-1}$}}
\def\ipb{\mbox{pb$^{-1}$}}
\newcommand{\meff}{\ensuremath{m_{\mathrm{eff}}}}

\markboth{G. Redlinger}
{Searches for supersymmetry with the ATLAS detector}

\catchline{}{}{}{}{}


\title{SEARCHES FOR SUPERSYMMETRY WITH THE ATLAS DETECTOR}

\author{\footnotesize G.Redlinger}

\address{Physics Department, Brookhaven National Laboratory, \\
  Upton, New York 11973, USA\\
  redlinger@bnl.gov}

\maketitle


\begin{abstract}
  This is a review of searches for supersymmetry (SUSY) with the ATLAS
  detector in proton-proton collisions at a center-of-mass energy of 7
  TeV at the Large Hadron Collider at CERN.  The review covers results
  that have been published, or submitted for publication, up to
  September 2012, many of which cover the full 7 TeV
  data-taking period.  No evidence for SUSY has been seen; some
  possibilities for future directions are discussed.

\end{abstract}

\ccode{PACS Nos.: 12.60.Jv, 13.85.Rm, 14.80.Ly}

\section{Introduction}	

One of the cornerstones of the physics program at the Large Hadron
Collider \cite{Evans:2008zzb} (LHC)  is the search for new phenomena
at the TeV scale, motivated by  
the gauge hierarchy
problem.
The recent discovery at the LHC
\cite{:2012gk,:2012gu}  of what may be the Higgs boson gives further
reality to the hierarchy problem.  Among the many possibilities for a solution,
supersymmetry (SUSY) is perhaps the most attractive; it is certainly
the most studied.  A sampling of recent reviews on the hierarchy
problem and SUSY  can be
found in the
references \cite{Feng:2009te,Peskin:2008nw,Djouadi:2005gj,Pape:2006ar,Barbier:2004ez,Chemtob:2004xr,Chung:2003fi,Martin:1997ns}.
Despite the lack of clear experimental evidence,\footnote{It could
  be argued that the closest exception might be the 
  measurement of the anomalous magnetic moment of
  the muon \cite{Stockinger:2006zn}.}
SUSY remains an attractive possibility.  In addition to stabilizing
the gauge hierarchy in the presence of radiative corrections, SUSY
provides an example of the unification of the gauge coupling
constants, a mechanism for radiative electroweak symmetry breaking,
a possible candidate for dark matter and, unique among models of new
phenomena, a 
framework for the unification of particle physics with gravity.

In the Minimal Supersymmetric Standard Model (MSSM), every Standard
Model fermion has a bosonic partner, and vice versa.  The Higgs sector
of the MSSM contains two Higgs doublets.  Gluinos $(\tilde{g})$ and
squarks $(\tilde{q})$ are the SUSY partners of gluons and squarks.
Charginos $(\tilde{\chi}_{i}^{\pm}, i=1,2)$ and neutralinos
$(\tilde{\chi}_{i}^{0}, i=1-4)$ (hereafter collectively referred to as
``gauginos'') are the mass eigenstates formed from
the linear superposition of the SUSY partners of the Higgs and
electroweak gauge bosons: higgsinos, winos and the bino.  The SUSY
partners of the charged leptons and neutrinos are collectively referred to as
sleptons $(\tilde{\ell}^{\pm})$ and sneutrinos $(\tilde{\nu})$,
respectively. R-parity conservation is
introduced in order to prevent the occurrence of
baryon and lepton number violating processes which are severely
constrained experimentally.  All Standard Model particles have
R-parity of 1, while the superpartners have R-parity of -1.  R-parity
conservation implies that SUSY particles are produced in pairs and
that the lightest SUSY particle (LSP) is stable.  In a large fraction
of the SUSY parameter space, the LSP is the weakly interacting
lightest neutralino, $\tilde{\chi}_{1}^{0}$, a potential  candidate
for dark matter; this gives rise to the classic SUSY signature of
missing transverse momentum (with magnitude denoted by \met).

If SUSY is to provide a natural solution to the hierarchy
problem, this implies bounds on the SUSY particle
masses \cite{Barbieri:1987fn,deCarlos:1993yy}.  The bounds depend on a
number of factors, not least of which are
the definition and amount of fine-tuning,
but general expectations (for a recent treatment, see
Ref. \cite{Papucci:2011wy})  in the MSSM
are that the higgsinos, top
squarks, the left-handed bottom squark and the gluino should not have
masses too far above the weak scale, while other SUSY particles could
in principle have masses beyond the reach of the LHC.
Rough numbers are 200, 600 and 900 GeV for
the higgsino, stop/sbottom and gluino masses,
respectively.\footnote{The recent ``Higgs''
discovery  changes this picture somewhat in the MSSM; the mixing between the
left- and right-handed
stop states needs to be large (see, for example, Ref. \cite{Carena:2011aa})
in order to accommodate a Higgs boson
mass in the observed range, resulting in a significant mass
splitting between the two stop mass eigenstates. Extensions to the
MSSM can evade these restrictions.}
Such a
configuration of SUSY particle masses sometimes goes by the name of
``Natural SUSY''.

This review summarizes the ATLAS searches for SUSY in 7 TeV
proton-proton collisions at the LHC.  Unless otherwise specified, the
searches are based on the complete dataset taken at the 7 TeV
center-of-mass energy, corresponding to an integrated luminosity of 4.7 \ifb.
The review covers results
that have been published, or submitted for publication, up to
September 2012.  Searches for the SUSY Higgs bosons will not be
covered. 
  
\section{The ATLAS detector}

The ATLAS detector 
\cite{Aad:2008zzm,Aad:2009wy} consists of a
tracking system (inner detector, ID)
surrounded by a
thin superconducting solenoid providing a 2~T magnetic field,
electromagnetic and hadronic
calorimeters and a muon spectrometer (MS).  The ID consists of pixel
and silicon microstrip detectors, surrounded by a straw-tube tracker
with transition radiation detection (transition
radiation tracker, TRT).  The electromagnetic calorimeter is a
lead liquid-argon (LAr) detector.  Hadronic calorimetry is based on two
different detector technologies, with scintillator-tiles or LAr 
as active media, and with either steel, copper, or tungsten as the
absorber material.
The MS is based on
three large superconducting toroid systems arranged with an eight-fold
azimuthal coil symmetry around the calorimeters, and three
stations of chambers for the trigger and  for precise position measurements.

\section{Modeling of Standard Model backgrounds and SUSY signal}

Samples of simulated events (with a detector simulation
\cite{Aad:2010ah} based on GEANT4 \cite{Agostinelli:2002hh})
are used for estimating the SUSY signal
acceptance, the detector efficiency, and for estimating many of the
Standard Model backgrounds.  For the latter, the simulation is
typically used in
conjunction with measurements in a background control region which
is designed to isolate a given 
process with the highest achievable purity and with kinematic
requirements 
as similar as possible to the signal region, while at the same time
minimizing contamination from potential signal. The simulation is
normalized to the event yield in the control region, and the
background in the signal region is estimated by extrapolating
via simulation the background level from the control region
to the signal region.  Unless stated otherwise, all searches described
in this review adopt this technique for the major background
processes.  For smaller backgrounds, the background is estimated
purely from simulation, using the best available theoretical cross
sections. 

For many SUSY searches, the most important backgrounds are \ttbar~and
$W$- or $Z$-boson production with multiple jets; these are produced
with multi-parton generators such as ALPGEN \cite{Mangano:2002ea}  or
SHERPA \cite{Gleisberg:2008ta}, with (in some cases)
up-to six additional partons in the
matrix element. The next-to-leading-order (NLO)
generators MC@NLO \cite{Frixione:2002ik} and POWHEG \cite{Powheg} are
also used for \ttbar~and single-top production.
Diboson\footnote{$WW$, $WZ$ and $ZZ$, where $Z$ implies also
  $\gamma^{\ast}$.}
production is
usually generated with HERWIG \cite{Corcella:2000bw}; SHERPA is used
when jet and/or photon emission are significant issues.
MADGRAPH5 \cite{Alwall:2011uj} is used for the production of \ttbar~in
association with $W$ or $Z$ bosons.  Parton shower and fragmentation
processes are simulated with
either HERWIG or PYTHIA \cite{Pythia}.

The cross section for the production of SUSY particles is
calculated in the MSSM at NLO precision in the strong coupling
constant, including the resummation of soft gluon emission at
next-to-leading-logarithmic (NLO+NLL)
accuracy, using PROSPINO and NLL-fast \cite{Beenakker:1996ch,Kulesza:2008jb,Kulesza:2009kq,
  Beenakker:2009ha,Beenakker:2011fu}.  The subsequent decay of these
SUSY particles depends on the SUSY breaking which has approximately
100 associated parameters.
As it is impossible to cover the entire parameter space by simulation,
several complementary approaches are taken
when estimating the sensitivity of the searches to SUSY signals.  In the
first approach, complete SUSY models are simulated; these models
typically impose boundary conditions at a high energy scale, reducing
the number of parameters to about five, and making it realistic to scan
the parameter space by brute force.  Examples studied in ATLAS are
MSUGRA/CMSSM, and minimal GMSB and AMSB models.  The computation of
the SUSY particle masses and branching ratios at the weak scale from
the high-scale parameters is performed with a number of publicly
available computational tools and is subject to non-negligible
theoretical uncertainty
\cite{Allanach:2003jw} which are not taken into account in ATLAS SUSY
limits.  In the second,
so-called ``simplified model'' \cite{Alwall:2008ag,Alves:2011wf},
approach, the SUSY decay cascades are simplified by setting the masses
of most SUSY particles to multi-TeV values, putting them out of range
of the LHC.  The decay cascades of the remaining particles
to the LSP, typically with zero or one
intermediate step, is characterized only by the masses of the
participating particles,\footnote{Branching ratios can be set by hand
if desired} allowing studies of the search sensitivity  to the SUSY
masses and decay kinematics.
The third approach is based on the
phenomenological MSSM (pMSSM) \cite{Djouadi:1998di} which reduces the
number of MSSM parameters to 22 by assuming the absence of new sources of
flavor changing neutral currents and CP violation.  By sampling a limited 
number of pMSSM parameters, the sensitivity of the searches to
more ``realistic'' configurations of SUSY particle masses and branching ratios
can  be assessed.  SUSY signal samples are typically generated with
either Herwig++ \cite{Bahr:2008pv} or MADGRAPH; the latter is usually
used (with an additional parton in the matrix element)
if initial-state radiation is important to the signal acceptance.

\section{Searches for gluinos and squarks}
\label{sec:sqgl}

For a fixed particle mass, squarks and gluinos
have the largest SUSY production cross sections
at the LHC.  They are thus prime candidates for the most
inclusive searches for SUSY.
The production of gluinos and squarks
($\tilde{u},\tilde{d},\tilde{s},\tilde{c}$) proceeds via $pp
\rightarrow \tilde{q}\tilde{q}, \tilde{q}\tilde{q}^{\ast},
\tilde{q}\tilde{g}, \tilde{g}\tilde{g}$. In simplified models with very heavy
squarks, gluinos decay via $\tilde{g} \rightarrow
q\overline{q}\tilde{\chi}_{i}^{0}$ or $qq'\tilde{\chi}_{i}^{\pm}$.  If gluinos
are very heavy, squarks decay via $\tilde{q}
\rightarrow q\tilde{\chi}_{i}^{0}$ or $q'\tilde{\chi}_{1}^{\pm}$.
Ignoring additional jets from initial- or final-state radiation, event
topologies with two, three and four jets are therefore expected for
$\tilde{q}\tilde{q}, \tilde{q}\tilde{g}$, and $\tilde{g}\tilde{g}$
production, respectively.  More complicated decay cascades lead
to larger numbers of jets in the final state.
When gauginos are produced in the decay
chain, leptons can be present via the decays $\tilde{\chi}_{1}^{\pm}
\rightarrow W^{(\ast)\pm} \tilde{\chi}_{1}^{0}$ or
$\tilde{\chi}_{2}^{0} \rightarrow Z^{(\ast)} \tilde{\chi}_{1}^{0}$.
The most inclusive searches for
SUSY are therefore
based on the presence of multiple jets, one or more leptons,
and missing transverse
momentum, where the latter arises (in
part) from the two LSP's in the event.  Useful observables include
\met~and $H_{\rm{T}}$,  defined as the scalar sum of the
transverse momenta of the jets and leptons\footnote{Unless otherwise
  specified, leptons identified in ATLAS will refer to electrons and
  muons.}
in the event.
The sum $H_{\rm{T}}+\met$, sometimes called the effective mass
(\meff), reflects the mass difference between the
initially-produced
SUSY particle and the LSP, and is approximately independent of the
details of the intermediate states in the decay cascade.

\subsection{Searches based on jets plus \met~with and without
  leptons}

ATLAS has two searches based on the jets plus \met~signature, vetoing
events containing electrons or muons in order to be orthogonal to
dedicated searches requiring leptons.  In the
first search \cite{:2012rz}, eleven different signal regions are defined, based
on different requirements on the jet multiplicity and on \meff.
Associated with each signal region are five background control
regions, each enhanced in one of the main SM backgrounds:
semi-leptonic \ttbar, $W$+jets, $Z$+jets, $\gamma$+jets (for estimating
$Z$+jets), and QCD multijets.  Transfer factors converting the
event yields in the control regions to predictions for the background
in the signal region (and cross-contamination of other control
regions) are obtained from simulation, except for the case
of QCD multijets where the factor comes from data.
The background in each signal region is determined with a
likelihood fit to the event yields in the control regions, together
with the transfer factors. The results
from the different signal regions are combined by selecting, for each
SUSY signal sample, the signal region that gives the best expected
limit.

\begin{figure}[ht]
\centerline{\psfig{file=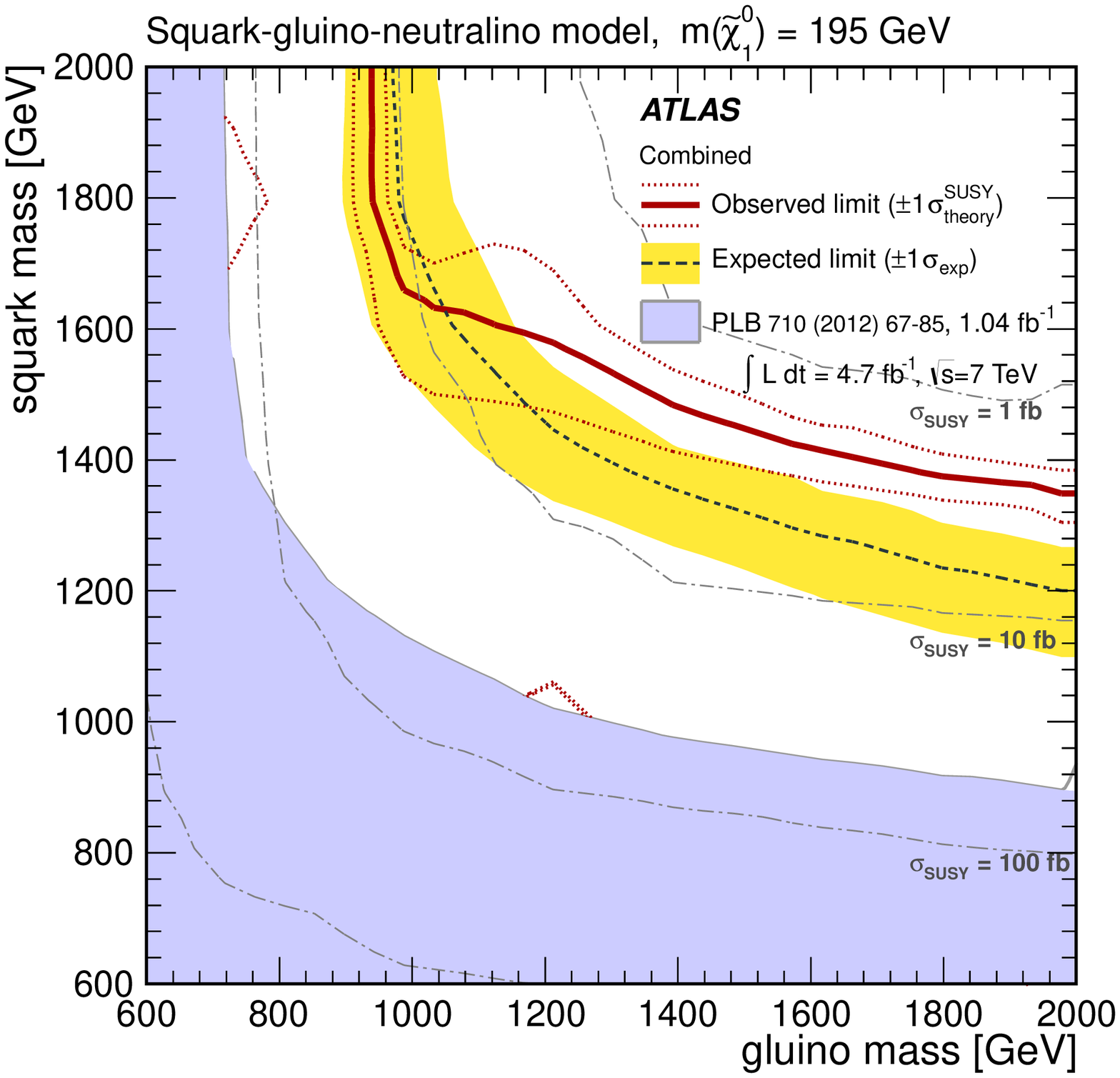,width=0.49\textwidth}\psfig{file=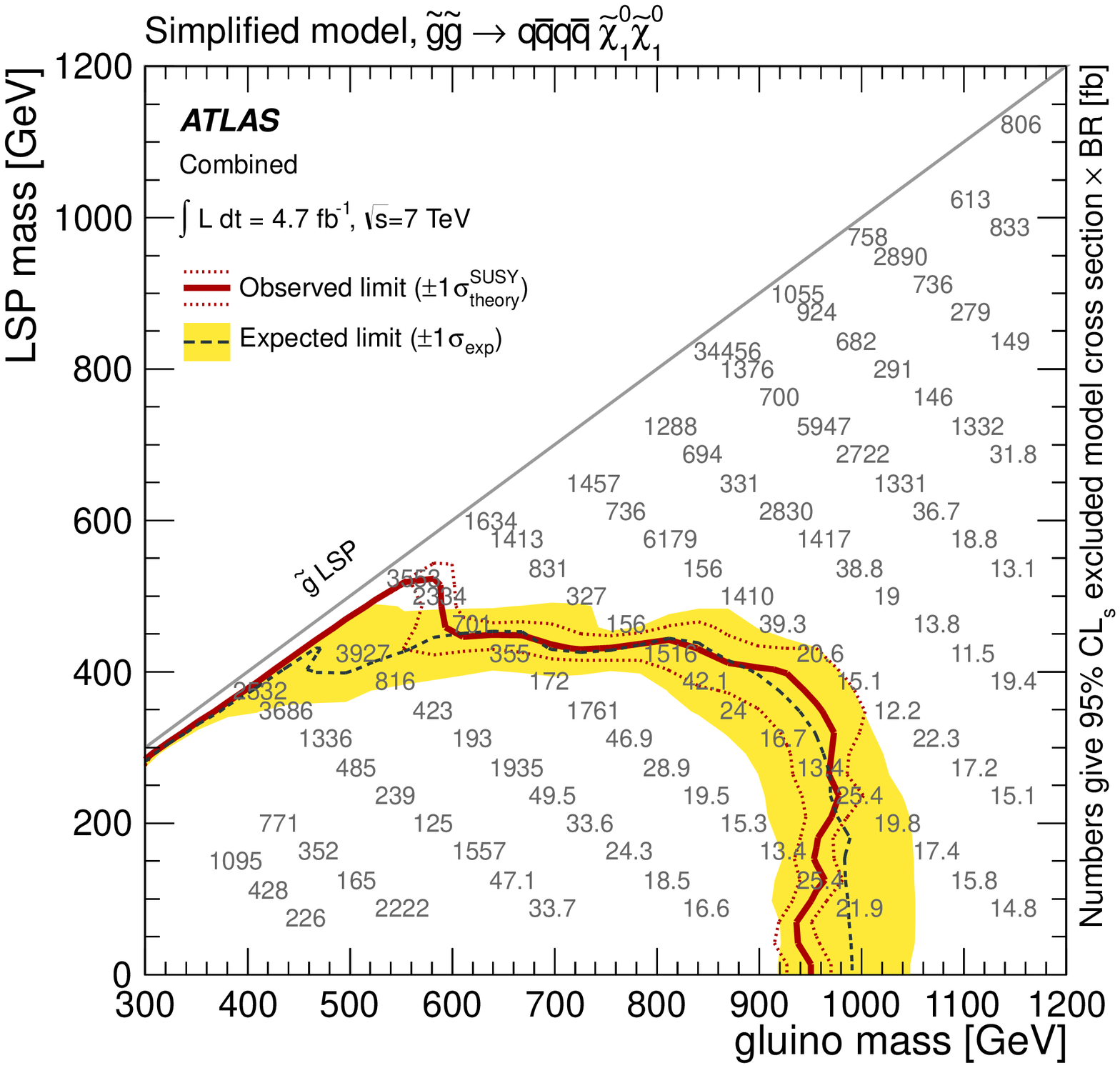,width=0.49\textwidth}}
\vspace*{8pt}
\caption{Left: 95\% CL exclusion limits in the plane of the squark mass
 versus gluino mass in a simplified MSSM model consisting of the
 gluino, squarks of the first and second generation, and the LSP,
 with direct decays of squarks and gluinos to jets and the LSP.  The
 LSP mass is set to 195 GeV.  Right: 95\% CL exclusion limits in a
 simplified model of gluino pair production with decoupled squarks;
 the 95\% CL upper limit on the cross section times branching ratio
 (in fb) is printed for each model point.
 For both plots, the band around the median expected
 limit shows the $\pm 1 \sigma$ variations, including all
  uncertainties except theoretical uncertainties on the signal. The
  dotted lines around the observed limit indicate the sensitivity to
  $\pm 1\sigma$ variations on these theoretical uncertainties.}

\protect\label{fig:sqgl}
\end{figure}

In a simplified
model consisting of the gluino, squarks of the first and second
generation and the LSP, gluinos (squarks)
with a mass below approximately 860 GeV (1320 GeV) are
excluded\footnote{All exclusion limits are quoted at 95\% confidence
  level.} as shown in Fig. \ref{fig:sqgl} (left);
 these results hold for LSP masses as high as approximately
400 GeV.  Limits are also derived in the MSUGRA/CMSSM model, and in
simplified models of $\tilde{g}\tilde{g}$ or $\tilde{q}\tilde{q}$
production followed either by direct decay to the LSP or via
an intermediate chargino.  In models with only gluino pair production
and direct decay to the LSP, gluinos with mass below about 550 GeV are
almost entirely excluded, independent of the LSP mass, as shown in
Fig. \ref{fig:sqgl} (right); however, no
limits can be set for gluino masses above about 600 GeV if the LSP
mass is above approximately 400 GeV.  Limits are generally weaker
in models where SUSY production is limited to squark production, due
primarily to the lower production cross section.
Limits in a variety of
``compressed SUSY'' models \cite{LeCompte:2011cn,LeCompte:2011fh} are
also derived.

The second search \cite{Aad:2012hm} is geared towards final states with
high jet multiplicities, ranging from $\ge 6$ to $\ge 9$ jets,
requiring the presence of \met~and vetoing on electrons and muons.
The search is performed in the variable $\met/\sqrt{H_{\rm{T}}}$.  The
background from QCD multijets dominates, followed by \ttbar.  The
former is estimated by taking advantage of the observation that
$\met/\sqrt{H_{\rm{T}}}$ is independent of jet multiplicity, thereby
allowing the shape of this quantity
to be determined in low jet multiplicity samples.
Other backgrounds are estimated using methods similar to the analysis
described previously.  Results are interpreted in the context of the
MSUGRA/CMSSM model and a simplified model of gluino pair production
where each gluino decays  only via $\tilde{g} \rightarrow
t\overline{t}\tilde{\chi}_{1}^{0}$.  Gluinos with mass below about 850
GeV are excluded in both scenarios.

Searches based on jets plus \met~and additional leptons 
\cite{:2012ms} reach generally similar conclusions in
MSUGRA/CMSSM as well as in simplified models with intermediate
charginos.  Additional interpretations include limits in GMSB and in
simplified models with two intermediate states, for example gluino
decay to a chargino (or $\tilde{\chi}_{2}^{0}$)
followed by decay to the LSP via intermediate
sleptons, sneutrinos or gauge bosons.  Searches involving
taus \cite{ATLAS:2012ag,Aad:2012rt} as
well as same-sign dileptons \cite{ATLAS:2012ai} ($e$
and $\mu$) have also been published, albeit for an integrated
luminosity of 2.05 \ifb.

\subsection{Searches for gluinos and squarks of the third generation}

Searches for gluinos and squarks of the third generation have been
performed in a number of channels, including jets plus \met~with high
jet multiplicity \cite{Aad:2012hm}, same-sign
dileptons \cite{ATLAS:2012ai}, and one or more $b$-tagged jets with and
without additional leptons \cite{ATLAS:2012ah}, the last two with an
integrated luminosity of 2.05 \ifb.  Such searches are particularly
well-motivated by the ``Natural SUSY'' scenario.  The most sensitive
search \cite{:2012pq} 
requires four or six jets of which three or more must be $b$-tagged,
no leptons and significant \met.   The requirement on
the $b$-jet multiplicity is effective in reducing the dominant
\ttbar~background.  Limits are derived in several simplified models
with gluino pair production followed by the decays
$\tilde{g}\rightarrow \tilde{b}b$ or $\tilde{g}\rightarrow \tilde{t}t$
where the bottom/top squark can be virtual.  The final state for each
gluino is the
same, whether or not the intermediate squark is real or virtual:
$b\overline{b}\tilde{\chi}_{1}^{0}$ or
$t\overline{t}\tilde{\chi}_{1}^{0}$.  For the models with real bottom
and top squarks, the LSP mass is fixed to 60 GeV, while for the models
with virtual squarks, the limits are provided in the plane of gluino
and LSP masses.  For a LSP mass of 60 GeV, gluinos with mass below about
1000 (820) GeV are excluded for bottom (top) squark masses up to about
870 (640) GeV.

\subsection{Searches with photons plus \met}

An inclusive search based on two or more photons and
\met~\cite{Aad:2012unknown} is motivated by gauge-mediated SUSY models in
which the next-to-lightest SUSY particle (NLSP) is a bino-like
neutralino decaying to the gravitino LSP via $\tilde{\chi}_{1}^{0}
\rightarrow \gamma \tilde{G}$.  The search makes very few requirements
other than two or more photons and \met; in particular, no jet
requirements are applied.  The main backgrounds ($\gamma\gamma$+X,
$\gamma$+jets+X, multijets, $W \rightarrow e\nu$+X and $\ttbar
\rightarrow e\nu$+X)  are estimated directly
from the data.  Based on a general gauge
mediation \cite{Meade:2008wd,Buican:2008ws,Ruderman:2011vv} model in which
the mass of the NLSP and either the squark or gluino are free parameters,
squarks (gluinos) with a mass below 870 (1070) GeV are excluded for
bino masses above 50 GeV.

\subsection{Searches for stable gluinos and squarks}

The possibility of a bound state containing a gluino or a squark,
the so-called R-hadron, was raised already in the earliest papers on MSSM
phenomenology  \cite{Farrar:1978xj}.  For more recent reviews of
stable massive particles, see
Ref. \cite{Fairbairn:2006gg,Raklev:2009mg}.
To date, ATLAS has published
results on R-hadrons only on data taken in 2010, corresponding to  an
integrated luminosity of 31-37 \ipb, although this was one of the
earliest SUSY publications from ATLAS.  Preliminary results with
the full 2011 dataset have been shown in conferences but will not be
discussed here.  The experimental signature for R-hadrons is
complicated  by the fact that they could have a significant
probability of undergoing hadronic
reactions in the detector material  \cite{Mackeprang:2009ad}.  Several
searches are therefore designed, utilizing different portions of the
ATLAS detector.

In the first approach \cite{Aad:2011yf}, the strategy
combines a search for heavily ionizing particles in the pixel detector
with a search for a slow-moving massive particle based on timing
information from the hadronic calorimeter.  This search is potentially
sensitive to R-hadrons even if they interact in the dense calorimeter
material.  Triggering is a challenge as R-hadron pair production could
be accompanied by very little other detector activity.  A \met~trigger
is therefore used to trigger on events where the production of R-hadrons is in
association with initial-state radiation in the $pp$ collision.  The
background, primarily from mis-measured muons, is estimated entirely
from the data, utilizing the lack of correlation between the particle
momentum, ionization and time-of-flight.  Stable sbottoms, stops and
gluinos with masses below 294, 309 and 562 GeV, respectively, are
excluded, using a conservative model for R-hadron scattering.

The second search \cite{Aad:2011hz} utilizes the
time-of-flight to the muon
spectrometer, with its superior timing resolution, as an additional
handle; this search is complementary to the first search in that it is
sensitive to R-hadrons that do not leave detectable signals in the
inner part of the detector, for example by being neutral at birth.
Backgrounds are estimated directly from the data as in the first
search.  Gluino R-hadrons with masses below 530 GeV are excluded.

In the
third search \cite{Aad:2012zn}, R-hadrons that have stopped
in the calorimeter material are searched for by looking for
calorimeter activity during periods in the LHC bunch structure without
$pp$ collisions.  This search complements the previous two searches
which are less sensitive to particles with $\beta \ll 1$.  The
dominant background from cosmic rays is measured from the data,
utilizing periods with long live time but small integrated luminosity.
Gluinos with mass between 200 and 341 GeV are excluded for lifetimes
between $10^{-5}$ and $10^{3}$ seconds, assuming a LSP mass of 100
GeV.

Although these searches cover a broad range of mechanisms for
R-hadron interactions, there remains the possibility that a R-hadron
remains neutral through the entire detector.  This case would be
covered by
the search for a monojet plus \met~signal, discussed in
Sec. \ref{sec:gaugino}. 

\section{Searches for third-generation squarks}

Searches for third-generation squarks are well-motivated by ``Natural
SUSY'' considerations.  As discussed in Sec. \ref{sec:sqgl}, if
gluinos are accessible at the LHC, it is likely that third-generation
squarks would first be seen in gluino decay cascades.  To cover the
possibility where the gluinos are beyond reach, searches for direct
production of third-generation squarks have also been made by ATLAS.  The
first dedicated search in ATLAS for stop production was performed with
2.05 \ifb~of data, motivated by a ``natural'' GMSB
scenario \cite{Asano:2010ut} with light stops and light higgsinos where
the higgsino NLSP decays to the gravitino via $\tilde{\chi}_{1}^{0}
\rightarrow Z \tilde{G}$ with a branching ratio ranging from 0.65 to
1, depending on the higgsino mass, and the stops decay via either
$\tilde{t}_{1}\rightarrow b \tilde{\chi}_{1}^{+}$ or (if kinematically
allowed) via $\tilde{t}_{1}\rightarrow t \tilde{\chi}_{1(2)}^{0}$.
The search requires two leptons consistent with coming from a
$Z$-boson, two or more jets, of which at least one must be $b$-tagged,
and significant \met.  Stops with mass below 240 GeV are excluded for
all NLSP masses (provided the NLSP mass is greater than the $Z$ boson
mass).  For NLSP masses between 115 and 230 GeV, the limit on the stop
increases to 310 GeV.

Less model-dependent searches for stops have been performed in a
number of channels.  The stop decay chains considered are $\tilde{t}_{1}
\rightarrow b \tilde{\chi}_{1}^{\pm} \rightarrow
b W^{(\ast)} \tilde{\chi}_{1}^{0}$ and, if kinematically allowed,
$\tilde{t}_{1} \rightarrow t \tilde{\chi}_{1}^{0}$.  The searches are
challenging due to the similarity of these final states to the
high background from \ttbar~production.
A search \cite{:2012tx} for stops lighter than the
top quark looks for evidence of the decay chain $\tilde{t}_{1}
\rightarrow b \tilde{\chi}_{1}^{\pm}$ in events with two opposite-sign
leptons, at least one jet, and significant \met.
Assuming a chargino mass of 106 GeV, identical to that assumed in a
previous analysis by CDF \cite{Aaltonen:2009sf}, top squarks with mass
below 130 GeV are excluded for LSP masses between 1 and 70 GeV.

Searches for stop with mass heavier than the top quark have been
performed in the decay mode $\tilde{t}_{1} \rightarrow t
\tilde{\chi}_{1}^{0}$.  The final state, consisting of two top quarks
and \met, has been searched for in the cases where both tops decay
hadronically \cite{:2012si} and where one top decays semi-leptonically 
\cite{:2012ar}.  In the all-hadronic analysis, six or more jets are
required, one or more of which are $b$-tagged, and
three of which are required to have a mass consistent with
that of the top quark.  Leptons are vetoed; kinematic
cuts are used to suppress the dominant background from semi-leptonic
\ttbar~production where the lepton is lost or mis-identified.  \met~is
used as the final discriminant.  Top squarks with mass between 370 and
465 GeV are excluded for a nearly massless LSP.  In the single-lepton
analysis, in addition to the presence of a lepton,
four or more jets are required, one or more of which are $b$-tagged.
The dominant background from di-leptonic \ttbar~(where one lepton is
lost or mis-identified) is suppressed by
requiring three jets to have a mass consistent with that of the top
quark and by requiring large transverse mass between the lepton
and \met.  Top squarks with mass between 230 and 440 GeV are excluded
for a nearly massless LSP.  Stop masses around 400 GeV are excluded
for LSP masses up to 125 GeV; for lower stop masses (but still greater
than 230 GeV), stops are excluded for all LSP masses
ranging up to approximately 50 GeV of the kinematic limit.

It should be noted that the above stop mass limits are weakened
if the stop decays with a mixture of branching ratios.  Scans in the
pMSSM parameter space 
in a region compatible with ``Natural SUSY'' indicate that a mixture
of at least two decay modes is very likely \cite{Cao:2012rz}.

Searches for direct production of bottom squarks have been published
with an integrated luminosity of 2.05 \ifb~\cite{Aad:2011cw}.  The
search assumes that the sbottom decays exclusively via $\tilde{b}_{1}
\rightarrow b \tilde{\chi}_{1}^{0}$ and looks for final states with
two high-\pt~ $b$-jets and significant \met.  Leptons
are vetoed.  The final discriminant
is the boost-corrected contransverse mass \cite{Polesello:2009rn}.
The major backgrounds are \ttbar, single-top and vector bosons
produced in association with $b$-jets.  For LSP masses below 60 GeV,
sbottom masses below 390 GeV are excluded.

\section{Searches for sleptons}

The search for directly produced, promptly decaying, sleptons is
challenging due to the small cross section and the large backgrounds
from diboson and \ttbar~backgrounds.  The ATLAS search \cite{:2012gg}
is based on the decay
chain $\tilde{\ell} \rightarrow \ell \tilde{\chi}_{1}^{0}$, and looks
for two opposite-sign, same-flavor leptons with significant \met~and
no jets in the event.  The final discriminating variable is
$m_{\rm{T2}}$~\cite{Barr:2003rg,Lester:1999tx} which falls off above
the $W$-boson mass for the background.   Left-handed selectrons and smuons
with masses between 85 and
195 GeV are excluded, assuming a LSP mass of 20 GeV; no limit is
possible for LSP masses above approximately 75 GeV.

Limits on sleptons with detector-scale lifetimes (or longer) can be
derived from the slow-particle searches described in
Sec. \ref{sec:sqgl}.  The MS-based search of Ref. \cite{Aad:2011hz}
excludes stable $\tilde{\tau}$s for masses below 136 GeV in a
particular GMSB model.  Directly-produced sleptons, which are expected
to be less model-dependent, are excluded below a mass of 110
GeV in the same model.

\section{Searches for charginos and neutralinos}
\label{sec:gaugino}

Direct pair production of gauginos at the LHC is
dominated by $\tilde{\chi}_{1}^{+}\tilde{\chi}_{2}^{0}$ and
$\tilde{\chi}_{1}^{+}\tilde{\chi}_{1}^{-}$ production.
Chargino pair production has been probed in a
search based on a dilepton plus \met~signature (in $ee$, $\mu\mu$ and
$e\mu$ channels)
and a veto on jets and $Z$-bosons \cite{:2012gg},
exploiting the decay $\tilde{\chi}_{1}^{+} \rightarrow
\ell^{+}\nu\tilde{\chi}_{1}^{0}$. 
The interpretation is performed in a simplified model
where the chargino decays to the LSP via an intermediate slepton or
sneutrino, i.e. $\tilde{\chi}_{1}^{+} \rightarrow \tilde{\ell}_{L} \nu
\rightarrow \ell \nu \tilde{\chi}_{1}^{0}$ or
$\tilde{\chi}_{1}^{+} \rightarrow \tilde{\nu} \ell
\rightarrow \ell \nu \tilde{\chi}_{1}^{0}$, where the mass of the
intermediate state is set halfway between those of the chargino and
LSP.  Using the signal region based on $m_{\rm{T2}}$ devised for the
prompt slepton search,  charginos with mass between 110 and 340 GeV
are excluded for a near-massless LSP.  For charginos between 110 and
approximately 300 GeV, the excluded region extends to all LSP masses
up to approximately 100 GeV below the chargino mass.
It is important to note that there is
no exclusion sensitivity for the case where the chargino decays via a
(real or virtual) $W$-boson, without intermediate sleptons/sneutrinos.

The search for associated chargino-neutralino production is performed
in both dilepton and trilepton plus \met~channels.  In the trilepton
analysis \cite{:2012ku} three signal regions are defined. Two of the regions
veto events containing either
$b$-jets or a lepton pair consistent with the $Z$-boson;  
in one of these two regions the transverse mass,
computed from the unpaired lepton and the \met~is required to be
past the endpoint for $W$-bosons, while for the other region, more
stringent \pt~requirements are applied instead to all three leptons.
The third signal region requires the presence of a $Z$-boson in the
final state and applies the transverse mass requirement. The final
discriminating observable is \met.  In the dilepton analysis
\cite{:2012gg}, the
signal region based on $m_{\rm{T2}}$ for the slepton/chargino search
is employed.  Results are interpreted in a pMSSM framework where the
strongly interacting SUSY particles are all decoupled and the gaugino
mass parameters ($M_{1}, M_{2}$ and $\mu$) are varied.  To boost the
leptonic branching ratios, a slepton is inserted halfway in between
the two lightest neutralino states.  The dilepton and trilepton
channels are combined, with the dilepton channel contributing
especially for higher values of $M_{1}$.  Results are also interpreted
in two simplified models of $\tilde{\chi}_{1}^{+}\tilde{\chi}_{2}^{0}$
production. In the first, the gaugino decays to the LSP
via an intermediate slepton or sneutrino.  In the second model, the
decay proceeds via $W$ and $Z$ bosons.  The trilepton analysis has
better sensitivity for these models.  In the model with intermediate
sleptons/sneutrinos, degenerate
$\tilde{\chi}_{1}^{+}, \tilde{\chi}_{2}^{0}$ with masses up to 500 GeV
are excluded for large mass differences from the LSP.  Small regions
of the parameter space are excluded as well
in the model with decays via gauge bosons.

Limits on directly produced gauginos have also been
derived in minimal GMSB scenarios in the diphoton plus \met~analysis
\cite{Aad:2012unknown}, assuming the SPS8 model \cite{Allanach:2002nj}.  A
lower limit of 196 TeV is set on the SPS8 breaking scale,
corresponding to lower
limits on $\tilde{\chi}_{1}^{\pm}$ and $\tilde{\chi}_{1}^{0}$ 
masses of approximately 530 and 280 GeV, respectively.

A search for unstable charginos in an AMSB-inspired scenario has been
performed with an integrated luminosity of 1 \ifb.  In these
scenarios, the lightest chargino and the LSP are nearly degenerate
such that the chargino decay proceeds via $\tilde{\chi}_{1}^{+}
\rightarrow \pi^{+} \tilde{\chi}_{1}^{0}$ where the pion has a
momentum of the order of 100 MeV.  The search looks for events in
which an isolated,
high-\pt~charged track ``disappears'' in the ATLAS tracking volume, the
low-momentum pion going unobserved.  The main backgrounds are from
charged hadrons interacting with the detector material and from poorly
reconstructed  low-\pt~tracks which scatter in the tracker.
Templates for the shape of the track \pt~distribution from these two
background sources are determined from control regions in the
data. Limits are derived from a fit to 
the track \pt~distribution in
data, using the above background templates and signal templates
derived from simulation.  Charginos with mass less than 92 GeV and a
lifetime between 0.5 and 2 ns are excluded in this scenario.

A search for a high-\pt~jet plus significant \met~would be sensitive
to scenarios of direct gaugino production where the gaugino decay products are
too soft to be detected, or more generally to production of weakly
interacting particles \cite{Beltran:2010ww,Goodman:2010ku} in association
with a jet from initial-state radiation.
ATLAS has published one search for the monojet signature
\cite{Aad:2011xw} with 33 \ipb~of data.  The result is interpreted in
terms of limits on a model of large extra dimensions; ATLAS provides
no dark matter interpretation.  The  subject of collider searches for
dark matter is beyond the scope of this review, but examples of
limits derived from LHC results can be found
in Ref. \cite{Rajaraman:2011wf,Fox:2011pm}. 

\section{Searches for R-parity violating SUSY}

R-parity conservation is imposed by hand in order to evade strong
experimental constraints on baryon and lepton number violation.
Although R-parity conservation has the feature  that the
LSP is stable and therefore a candidate for dark matter, R-parity
violation (RPV) is perhaps equally well-motivated theoretically.  The
first consequence of RPV is an increase in the number of parameters,
making signal modeling even more challenging.  Other consequences are
that SUSY particles may be singly produced in
$pp$ collisions and that the LSP can decay.  The decay of the LSP
implies lower \met, thereby evading many of the searches described
above, but resulting in a higher
multiplicity of final state objects such as jets and leptons,
typically with a resonant structure.  

A few searches have been performed in ATLAS with an explicit RPV SUSY
interpretation.  A search for a high-mass, opposite-sign $e\mu$ resonance
\cite{Aad:2011qr}, with an integrated luminosity of
1 \ifb, is sensitive to the process $d\overline{d} \rightarrow
\tilde{\nu}_{\tau} \rightarrow e \mu$.  A limit on the cross section
times branching ratio $\sigma(pp \rightarrow \tilde{\nu}_{\tau})
\times \rm{BR}(\tilde{\nu}_{\tau} \rightarrow e\mu$) is set as a
function of the $\tilde{\nu}_{\tau}$ mass, with a value of 4.5 fb for
a mass of 1 TeV.  A search employing the same final-state but with a
non-resonant signature \cite{Aad:2012yw} is sensitive to the
$t$-channel exchange of a lightest R-parity violating
up-type squark, assumed to be the top squark.  From an analysis of
2.1 \ifb~of data, a limit on the cross
section $\sigma(pp \rightarrow e\mu)$ via t-channel top squark
exchange is derived as a function of the $\tilde{t}$ mass, with a
value of 30 fb for a mass of 1 TeV.

The inclusive search for SUSY in the lepton plus jets plus
\met~channel \cite{ATLAS:2011ad} with 1 \ifb~of data provides an
interpretation in the framework of a model where bilinear RPV
couplings are embedded in a MSUGRA/CMSSM SUSY production model. For a
chosen set of MSUGRA/CMSSM parameters, the bilinear RPV parameters are
fixed by a fit to neutrino oscillation data.
Limits are then derived as a function of MSUGRA/CMSSM parameters.

A search for new, heavy particles that decay at a significant distance
from their production point into a final state consisting of charged
hadrons in association with a muon has been performed with 33 \ipb~of
data \cite{Aad:2011zb}.
The search is sensitive, for example, to a neutralino LSP
decaying via $\tilde{\chi}_{1}^{0} \rightarrow \mu\tilde{\mu}
\rightarrow \mu qq'$.
The vertex position is required to be at a radius greater than 4
mm and within the fiducial volume of the pixel detector ($r < 180$ mm
and $|z| < 300$ mm).  After vetoing on vertices emerging from locations of known
detector material, signal is distinguished from background using
the number of tracks in the vertex and the vertex mass.
The main background source is a
low-mass vertex arising from a particle interaction with air, randomly matched
with another track, possibly from a different primary interaction.
The expected
number of background events in the signal region,
obtained from simulation validated against data,
is very low ($< 0.03$ events), and no events are observed.
A limit on the production cross
section times branching ratio is obtained as a function of neutralino
decay distance.

A search for resonances in 4-jet final states has been performed with
34 \ipb~of data.  A unique pairing of the four highest-\pt~jets
is defined for each event
by minimizing the pair-wise spatial separation, and the dijet mass
distribution is examined for signs of a resonance above background.
The shape of the background mass distribution is obtained from the
data by using two largely uncorrelated variables: the scattering angle
of the reconstructed particle in the rest frame of the 4-jet system,
and the relative mass difference between the two jet pairs.  The
results are interpreted in terms of the production of two
scalar gluons, the supersymmetric
partners of Dirac gluinos
\cite{Choi:2008ub,Kribs:2007ac,Plehn:2008ae},
each of which decay to two
gluons.  Scalar gluons with masses between 100 and 185 GeV are excluded.
Although the process of a scalar gluon decaying to two gluons is
R-parity conserving, the search for resonances in 4-jet final states
is sensitive to RPV decays as discussed, for example, in 
Ref. \cite{Ruderman:2012jd}.

Many other searches have been performed in ATLAS for physics beyond
the Standard Model.  Such searches are often sensitive to RPV SUSY,
although no such interpretation is provided by ATLAS.  A comprehensive
discussion of RPV limits from the re-interpretations of ATLAS results is
beyond the scope of this review.  Some recent examples in the
literature include Ref. \cite{Dreiner:2012mn}  and \cite{Allanach:2012vj}.

\section{Conclusion}

The discovery at the LHC of what may be the Higgs boson puts the 
gauge hierarchy problem front and center. If SUSY is to be a solution to the
hierarchy problem, limits on the masses of SUSY particles are
implied.  Such expectations are starting to be challenged by searches
at the LHC.

This review has covered the searches for SUSY conducted by ATLAS at
the LHC with a center-of-mass energy of 7 TeV.  The review is limited
to results in publication or submitted for publication up to
September 2012.  Many analyses covering the entire 7 TeV data-taking
period have been summarized here.  Preliminary results, updating many
of the remaining 
SUSY searches to the full 7 TeV dataset, have been shown in
conferences; first preliminary results from the 8 TeV data-taking are
also available.\footnote{The full list of ATLAS results can be found at
  http://twiki.cern.ch/twiki/bin/view/AtlasPublic} No evidence for
SUSY has been seen.

The most serious challenge to weak-scale SUSY comes from the searches
for gluinos, with and without lighter squarks.  A large variety of
searches for gluinos are reaching lower mass limits of 850 GeV or higher.
Some of the traditional ways out, such as compressed decay spectra, or
multi-step decay chains, are starting to get restricted.
The situation is less dire for the other SUSY particles required at the
weak scale: third-generation squarks and gauginos.
Searches for direct production of bottom and top squarks (assuming
gluinos are just out of reach) are entering
interesting territory, but limits degrade quickly when assumptions on
decay branching ratios are relaxed.  There also remains a challenging
window where the stop mass is close to the top mass.  The most stringent limits
on direct gaugino production are obtained only by boosting
the leptonic content by assuming intermediate sleptons in the decay
chains.  Limits are still very weak (or non-existent)
in the case where the decay is only via gauge bosons.

Several other ``escape routes'' have been discussed in the
literature.  One is to lower the production cross section of the
colored superpartners by assuming that the gluino acquires a large
Dirac mass; for a recent discussion, see Ref. \cite{Kribs:2012gx}.
Another is the possibility of ``stealth SUSY''; see
Ref. \cite{Fan:2012jf} for a survey of ``stealth'' models.
And
finally there is the possibility of R-parity violation, which weakens
the \met~signature on which so many SUSY searches rely, and which
comes with an even larger parameter space.  Many
searches have been performed in ATLAS for physics beyond the Standard
Model.  Only some have been interpreted by ATLAS in the context of
RPV SUSY models.  For a large number of searches, ATLAS provides
auxiliary information in HEPdata \cite{HEPdata} to ease the
re-interpretation of ATLAS results; this is an ongoing dialogue
with the community which can be expected to evolve as the needs become
clearer.  

ATLAS is currently taking data at a center-of-mass energy of 8 TeV.
The size of the dataset is expected to be on the order of 20 \ifb, and
further inroads into the territory of weak-scale SUSY can be
expected.  The rapidly evolving implications from the Higgs sector
may influence future SUSY search strategies.  If the newly discovered
particle turns out to be the Standard Model Higgs boson,
the complete exploration of the relevance of SUSY
to the hierarchy problem
will require the full LHC program at 14 TeV.

\section{Acknowledgements}

The author is grateful to D.\ Milstead,
A.\ Parker and A.\ Sfyrla for their review of this manuscript. This
work was supported in part under US Department of Energy contract
DE-AC02-98CH10886. 

\bibliographystyle{ws-ijmpa}
\bibliography{atlas_susy}

\begin{thebibliography}{10}

\bibitem{Evans:2008zzb}
L.~Evans and P.~Bryant~(editors), {\em JINST} {\bf 3},   S08001  (2008).

\bibitem{:2012gk}
 ATLAS Collaboration (G.~Aad {\em et~al.}), {\em Phys.Lett.} {\bf B716},  ~1
  (2012).

\bibitem{:2012gu}
 CMS Collaboration (S.~Chatrchyan {\em et~al.}), {\em Phys.Lett.} {\bf B716},
  ~30  (2012).

\bibitem{Feng:2009te}
J.~L. Feng, J.-F. Grivaz and J.~Nachtman, {\em Rev.Mod.Phys.} {\bf 82},   699
  (2010).

\bibitem{Peskin:2008nw}
M.~E. Peskin  (2008), \href{http://arxiv.org/abs/0801.1928}{{\ttfamily
  arXiv:0801.1928 [hep-ph]}}.

\bibitem{Djouadi:2005gj}
A.~Djouadi, {\em Phys.Rept.} {\bf 459},  ~1  (2008).

\bibitem{Pape:2006ar}
L.~Pape and D.~Treille, {\em Rept.Prog.Phys.} {\bf 69},   2843  (2006).

\bibitem{Barbier:2004ez}
R.~Barbier {\em et~al.}, {\em Phys.Rept.} {\bf 420},  ~1  (2005).

\bibitem{Chemtob:2004xr}
M.~Chemtob, {\em Prog.Part.Nucl.Phys.} {\bf 54},  ~71  (2005).

\bibitem{Chung:2003fi}
D.~Chung, L.~Everett, G.~Kane, S.~King, J.~D. Lykken {\em et~al.}, {\em
  Phys.Rept.} {\bf 407},  ~1  (2005).

\bibitem{Martin:1997ns}
S.~P. Martin  (1997), \href{http://arxiv.org/abs/hep-ph/9709356}{{\ttfamily
  arXiv:hep-ph/9709356 [hep-ph]}}.

\bibitem{Stockinger:2006zn}
D.~Stockinger, {\em J.Phys.G} {\bf G34},   R45  (2007).

\bibitem{Barbieri:1987fn}
R.~Barbieri and G.~Giudice, {\em Nucl.Phys.} {\bf B306},  ~63  (1988).

\bibitem{deCarlos:1993yy}
B.~de~Carlos and J.~Casas, {\em Phys.Lett.} {\bf B309},   320  (1993).

\bibitem{Papucci:2011wy}
M.~Papucci, J.~T. Ruderman and A.~Weiler  (2011),
  \href{http://arxiv.org/abs/1110.6926}{{\ttfamily arXiv:1110.6926 [hep-ph]}}.

\bibitem{Carena:2011aa}
M.~Carena, S.~Gori, N.~R. Shah and C.~E. Wagner, {\em JHEP} {\bf 1203},   014
  (2012).

\bibitem{Aad:2008zzm}
 ATLAS Collaboration (G.~Aad {\em et~al.}), {\em JINST} {\bf 3},   S08003
  (2008).

\bibitem{Aad:2009wy}
 ATLAS Collaboration (G.~Aad {\em et~al.})
  \href{http://arxiv.org/abs/0901.0512}{{\ttfamily arXiv:0901.0512 [hep-ex]}}.

\bibitem{Aad:2010ah}
 ATLAS Collaboration (G.~Aad {\em et~al.}), {\em Eur.Phys.J.} {\bf C70},   823
  (2010).

\bibitem{Agostinelli:2002hh}
 GEANT4 Collaboration (S.~Agostinelli {\em et~al.}), {\em Nucl. Instrum. Meth.}
  {\bf A506},   250  (2003).

\bibitem{Mangano:2002ea}
M.~L. Mangano, M.~Moretti, F.~Piccinini, R.~Pittau and A.~D. Polosa, {\em JHEP}
  {\bf 0307},   001  (2003).

\bibitem{Gleisberg:2008ta}
T.~Gleisberg {\em et~al.}, {\em JHEP} {\bf 0902},   007  (2009).

\bibitem{Frixione:2002ik}
S.~Frixione and B.~R. Webber, {\em JHEP} {\bf 0206},   029  (2002).

\bibitem{Powheg}
S.~Frixione, P.~Nason and C.~Oleari, {\em JHEP} {\bf 11},   070  (2007).

\bibitem{Corcella:2000bw}
G.~Corcella {\em et~al.}, {\em JHEP} {\bf 0101},   010  (2001).

\bibitem{Alwall:2011uj}
J.~Alwall, M.~Herquet, F.~Maltoni, O.~Mattelaer and T.~Stelzer, {\em JHEP} {\bf
  1106},   128  (2011).

\bibitem{Pythia}
T.~Sj{\"{o}}strand, S.~Mrenna and P.~Skands, {\em JHEP} {\bf 05},   026
  (2006).

\bibitem{Beenakker:1996ch}
W.~Beenakker, R.~Hopker, M.~Spira and P.~Zerwas, {\em Nucl. Phys.} {\bf B492},
  ~51  (1997).

\bibitem{Kulesza:2008jb}
A.~Kulesza and L.~Motyka, {\em Phys. Rev. Lett.} {\bf 102},   111802  (2009).

\bibitem{Kulesza:2009kq}
A.~Kulesza and L.~Motyka, {\em Phys. Rev.} {\bf D80},   095004  (2009).

\bibitem{Beenakker:2009ha}
W.~Beenakker {\em et~al.}, {\em JHEP} {\bf 0912},   041  (2009).

\bibitem{Beenakker:2011fu}
W.~Beenakker, S.~Brensing, M.~Kramer, A.~Kulesza, E.~Laenen {\em et~al.}, {\em
  Int. J. Mod. Phys.} {\bf A26},   2637  (2011).

\bibitem{Allanach:2003jw}
B.~Allanach, S.~Kraml and W.~Porod, {\em JHEP} {\bf 0303},   016  (2003).

\bibitem{Alwall:2008ag}
J.~Alwall, P.~Schuster and N.~Toro, {\em Phys. Rev.} {\bf D79},   075020
  (2009).

\bibitem{Alves:2011wf}
 LHC New Physics Working Group Collaboration (D.~Alves {\em et~al.})  (2011),
  \href{http://arxiv.org/abs/1105.2838}{{\ttfamily arXiv:1105.2838 [hep-ph]}}.

\bibitem{Djouadi:1998di}
 MSSM Working Group Collaboration (A.~Djouadi {\em et~al.})  (1998),
  \href{http://arxiv.org/abs/hep-ph/9901246}{{\ttfamily arXiv:hep-ph/9901246
  [hep-ph]}}.

\bibitem{Bahr:2008pv}
M.~Bahr {\em et~al.}, {\em Eur. Phys. J.} {\bf C58},   639  (2008).

\bibitem{:2012rz}
 ATLAS Collaboration (G.~Aad {\em et~al.})  (2012),
  \href{http://arxiv.org/abs/1208.0949}{{\ttfamily arXiv:1208.0949 [hep-ex]}}.

\bibitem{LeCompte:2011cn}
T.~J. LeCompte and S.~P. Martin, {\em Phys.Rev.} {\bf D84},   015004  (2011).

\bibitem{LeCompte:2011fh}
T.~J. LeCompte and S.~P. Martin, {\em Phys.Rev.} {\bf D85},   035023  (2012).

\bibitem{Aad:2012hm}
 ATLAS Collaboration (G.~Aad {\em et~al.}), {\em JHEP} {\bf 1207},   167
  (2012).

\bibitem{:2012ms}
 ATLAS Collaboration (G.~Aad {\em et~al.})  (2012),
  \href{http://arxiv.org/abs/1208.4688}{{\ttfamily arXiv:1208.4688 [hep-ex]}}.

\bibitem{ATLAS:2012ag}
 ATLAS Collaboration (G.~Aad {\em et~al.})  (2012),
  \href{http://arxiv.org/abs/1203.6580}{{\ttfamily arXiv:1203.6580 [hep-ex]}}.

\bibitem{Aad:2012rt}
 ATLAS Collaboration (G.~Aad {\em et~al.})  (2012),
  \href{http://arxiv.org/abs/1204.3852}{{\ttfamily arXiv:1204.3852 [hep-ex]}}.

\bibitem{ATLAS:2012ai}
 ATLAS Collaboration (G.~Aad {\em et~al.}), {\em Phys.Rev.Lett.} {\bf 108},
  241802  (2012).

\bibitem{ATLAS:2012ah}
 ATLAS Collaboration (G.~Aad {\em et~al.}), {\em Phys.Rev.} {\bf D85},   112006
   (2012).

\bibitem{:2012pq}
 ATLAS Collaboration (G.~Aad {\em et~al.})  (2012),
  \href{http://arxiv.org/abs/1207.4686}{{\ttfamily arXiv:1207.4686 [hep-ex]}}.

\bibitem{Aad:2012unknown}
 ATLAS Collaboration (G.~Aad {\em et~al.})  (2012),
  \href{http://arxiv.org/abs/1209.0753}{{\ttfamily arXiv:1209.0753 [hep-ex]}}.

\bibitem{Meade:2008wd}
P.~Meade, N.~Seiberg and D.~Shih, {\em Prog.Theor.Phys.Suppl.} {\bf 177},   143
   (2009).

\bibitem{Buican:2008ws}
M.~Buican, P.~Meade, N.~Seiberg and D.~Shih, {\em JHEP} {\bf 0903},   016
  (2009).

\bibitem{Ruderman:2011vv}
J.~T. Ruderman and D.~Shih, {\em JHEP} {\bf 1208},   159  (2012).

\bibitem{Farrar:1978xj}
G.~R. Farrar and P.~Fayet, {\em Phys. Lett.} {\bf B76},   575  (1978).

\bibitem{Fairbairn:2006gg}
M.~Fairbairn {\em et~al.}, {\em Phys.Rept.} {\bf 438},  ~1  (2007).

\bibitem{Raklev:2009mg}
A.~R. Raklev, {\em Mod.Phys.Lett.} {\bf A24},   1955  (2009).

\bibitem{Mackeprang:2009ad}
R.~Mackeprang and D.~Milstead, {\em Eur.Phys.J.} {\bf C66},   493  (2010).

\bibitem{Aad:2011yf}
 ATLAS Collaboration (G.~Aad {\em et~al.}), {\em Phys.Lett.} {\bf B701},  ~1
  (2011).

\bibitem{Aad:2011hz}
 ATLAS Collaboration (G.~Aad {\em et~al.}), {\em Phys.Lett.} {\bf B703},   428
  (2011).

\bibitem{Aad:2012zn}
 ATLAS Collaboration (G.~Aad {\em et~al.}), {\em Eur.Phys.J.} {\bf C72},   1965
   (2012).

\bibitem{Asano:2010ut}
M.~Asano, H.~D. Kim, R.~Kitano and Y.~Shimizu, {\em JHEP} {\bf 1012},   019
  (2010).

\bibitem{:2012tx}
 ATLAS Collaboration (G.~Aad {\em et~al.})  (2012),
  \href{http://arxiv.org/abs/1208.4305}{{\ttfamily arXiv:1208.4305 [hep-ex]}}.

\bibitem{Aaltonen:2009sf}
 CDF Collaboration (T.~Aaltonen {\em et~al.}), {\em Phys.Rev.Lett.} {\bf 104},
   251801  (2010).

\bibitem{:2012si}
 ATLAS Collaboration (G.~Aad {\em et~al.})  (2012),
  \href{http://arxiv.org/abs/1208.1447}{{\ttfamily arXiv:1208.1447 [hep-ex]}}.

\bibitem{:2012ar}
 ATLAS Collaboration (G.~Aad {\em et~al.})  (2012),
  \href{http://arxiv.org/abs/1208.2590}{{\ttfamily arXiv:1208.2590 [hep-ex]}}.

\bibitem{Cao:2012rz}
J.~Cao, C.~Han, L.~Wu, J.~M. Yang and Y.~Zhang  (2012),
  \href{http://arxiv.org/abs/1206.3865}{{\ttfamily arXiv:1206.3865 [hep-ph]}}.

\bibitem{Aad:2011cw}
 ATLAS Collaboration (G.~Aad {\em et~al.}), {\em Phys.Rev.Lett.} {\bf 108},
  181802  (2012).

\bibitem{Polesello:2009rn}
G.~Polesello and D.~R. Tovey, {\em JHEP} {\bf 1003},   030  (2010).

\bibitem{:2012gg}
 ATLAS Collaboration (G.~Aad {\em et~al.})  (2012),
  \href{http://arxiv.org/abs/1208.2884}{{\ttfamily arXiv:1208.2884 [hep-ex]}}.

\bibitem{Barr:2003rg}
A.~Barr, C.~Lester and P.~Stephens, {\em J.Phys.G} {\bf G29},   2343  (2003).

\bibitem{Lester:1999tx}
C.~Lester and D.~Summers, {\em Phys.Lett.} {\bf B463},  ~99  (1999).

\bibitem{:2012ku}
 ATLAS Collaboration (G.~Aad {\em et~al.})  (2012),
  \href{http://arxiv.org/abs/1208.3144}{{\ttfamily arXiv:1208.3144 [hep-ex]}}.

\bibitem{Allanach:2002nj}
B.~Allanach {\em et~al.}, {\em Eur.Phys.J.} {\bf C25},   113  (2002).

\bibitem{Beltran:2010ww}
M.~Beltran, D.~Hooper, E.~W. Kolb, Z.~A. Krusberg and T.~M. Tait, {\em JHEP}
  {\bf 1009},   037  (2010).

\bibitem{Goodman:2010ku}
J.~Goodman {\em et~al.}, {\em Phys.Rev.} {\bf D82},   116010  (2010).

\bibitem{Aad:2011xw}
 ATLAS Collaboration (G.~Aad {\em et~al.}), {\em Phys.Lett.} {\bf B705},   294
  (2011).

\bibitem{Rajaraman:2011wf}
A.~Rajaraman, W.~Shepherd, T.~M. Tait and A.~M. Wijangco, {\em Phys.Rev.} {\bf
  D84},   095013  (2011).

\bibitem{Fox:2011pm}
P.~J. Fox, R.~Harnik, J.~Kopp and Y.~Tsai, {\em Phys.Rev.} {\bf D85},   056011
  (2012).

\bibitem{Aad:2011qr}
 ATLAS Collaboration (G.~Aad {\em et~al.}), {\em Eur.Phys.J.} {\bf C71},   1809
   (2011).

\bibitem{Aad:2012yw}
 ATLAS Collaboration (G.~Aad {\em et~al.}), {\em Eur.Phys.J.} {\bf C72},   2040
   (2012).

\bibitem{ATLAS:2011ad}
 ATLAS Collaboration (G.~Aad {\em et~al.}), {\em Phys.Rev.} {\bf D85},   012006
   (2012).

\bibitem{Aad:2011zb}
 ATLAS Collaboration (G.~Aad {\em et~al.}), {\em Phys.Lett.} {\bf B707},   478
  (2012).

\bibitem{Choi:2008ub}
S.~Choi {\em et~al.}, {\em Phys.Lett.} {\bf B672},   246  (2009).

\bibitem{Kribs:2007ac}
G.~D. Kribs, E.~Poppitz and N.~Weiner, {\em Phys.Rev.} {\bf D78},   055010
  (2008).

\bibitem{Plehn:2008ae}
T.~Plehn and T.~M. Tait, {\em J.Phys.G} {\bf G36},   075001  (2009).

\bibitem{Ruderman:2012jd}
J.~T. Ruderman, T.~R. Slatyer and N.~Weiner  (2012),
  \href{http://arxiv.org/abs/1207.5787}{{\ttfamily arXiv:1207.5787 [hep-ph]}}.

\bibitem{Dreiner:2012mn}
H.~Dreiner and T.~Stefaniak  (2012),
  \href{http://arxiv.org/abs/1201.5014}{{\ttfamily arXiv:1201.5014 [hep-ph]}}.

\bibitem{Allanach:2012vj}
B.~Allanach and B.~Gripaios, {\em JHEP} {\bf 1205},   062  (2012).

\bibitem{Kribs:2012gx}
G.~D. Kribs and A.~Martin, {\em Phys.Rev.} {\bf D85},   115014  (2012).

\bibitem{Fan:2012jf}
J.~Fan, M.~Reece and J.~T. Ruderman, {\em JHEP} {\bf 1207},   196  (2012).

\bibitem{HEPdata}
\url{http://hepdata.cedar.ac.uk}.

\end{thebibliography}

\end{document}